\documentstyle[aps,pre,epsfig,psfig,eqsecnum]{revtex}
\def\eps{\varepsilon}
\def\k{{\bf k}}
\def\q{{\bf q}}
\def\bfr{{\bf r}}
\def\Dm{\widetilde{\cal D}_{\mu}}

\begin{document}
\title{Anomalous scaling of a passive scalar advected by the turbulent
velocity field with finite correlation time: Two-loop approximation}
\author{L.\,Ts.\,Adzhemyan$^{1}$,  N.\,V.\,Antonov$^{1}$, and
J.\,Honkonen$^{2}$}
\address{$^{1}$ Department of Theoretical Physics, St.~Petersburg University,
Ulyanovskaya 1, St.~Petersburg---Petrodvorez, 198504, Russia \\
$^{2}$ Theory~Division, Department~of~Physical Sciences,
P.O.~Box~64, FIN-00014 University~of~Helsinki, Finland} \draft
\date{24 December 2001}
\maketitle
\begin{abstract}
The renormalization group and operator product expansion are
applied to the model of a passive scalar quantity advected by the
Gaussian self-similar velocity field with finite, and not small,
correlation time. The inertial-range energy spectrum of the
velocity is chosen in the form $E(k)\propto k^{1-2\eps}$, and the
correlation time at the wavenumber $k$ scales as $k^{-2+\eta}$.
Inertial-range anomalous scaling for the structure functions and
other correlation functions emerges as a consequence of the
existence in the model of composite operators with negative
scaling dimensions, identified with anomalous exponents. For
$\eta>\eps$, these exponents are the same as in the rapid-change
limit of the model; for $\eta<\eps$, they are the same as in the
limit of a time-independent (quenched) velocity field. For
$\eps=\eta$ (local turnover exponent), the anomalous exponents are
nonuniversal through the dependence on a dimensionless parameter,
the ratio of the velocity correlation time and the scalar turnover
time. The universality reveals itself, however, only in the second
order of the $\eps$ expansion, and the exponents are derived to
order $O(\eps^{2})$, including anisotropic contributions. It is
shown that, for moderate $n$, the order of the structure function,
and $d$, the space dimensionality, finite correlation time
enhances the intermittency in comparison with the both limits: the
rapid-change and quenched ones. The situation changes when $n$
and/or $d$ become large enough: the correction to the rapid-change
limit due to the finite correlation time is positive (that is, the
anomalous scaling is suppressed), it is maximal for the quenched
limit and monotonically decreases as the correlation time tends to
zero.
\end{abstract}

\section{Introduction} \label{sec:Intro}

In recent years, considerable progress has been achieved in the
understanding of intermittency and anomalous scaling of fluid
turbulence. The crucial role in these studies was played by a
simple model of a passive scalar quantity advected by a random
Gaussian field, white in time and self-similar in space, the
so-called Kraichnan's rapid-change model \cite{Kraich1}. There,
for the first time the existence of anomalous scaling was
established on the basis of a microscopic model \cite{Kraich2} and
the corresponding anomalous exponents were calculated within
controlled approximations \cite{GK,Falk1,Pumir,Siggia} and a
systematic perturbation expansion in a formal small parameter
\cite{RG}. Detailed review of the recent theoretical research on
the passive scalar problem and the bibliography can be found in
Ref.~\cite{FGV}.

Within the approach developed in Refs.
\cite{GK,Falk1,Pumir,Siggia}, nontrivial anomalous exponents are
related to ``zero modes,"  that is, homogeneous solutions of the
closed exact differential equations satisfied by the equal-time
correlation functions. In this sense, the rapid-change model
appears ``exactly solvable.''

In a wider context, zero modes can be interpreted as statistical
conservation laws of the particle dynamics \cite{slow}. The
concept of statistical conservation laws appears rather general,
being also confirmed in numerical simulations by Refs.
\cite{CV,SCL}, where the passive advection in the two-dimensional
Navier--Stokes velocity field \cite{CV} and a shell model of a
passive scalar \cite{SCL} were studied. This observation is rather
intriguing because in those models no closed equations for
equal-time quantities can be derived due to the fact that the
advecting velocity has a finite correlation time (for a passive
field advected by a velocity with given statistics, closed
equations can be derived only for different-time correlation
functions, and they involve infinite diagrammatic series).

One may thus conclude that breaking the artificial assumption of
the time decorrelation of the velocity field is the crucial point
\cite{CV,SCL}.

An important issue related to the effects of the finite
correlation time is the universality of the anomalous exponents.
It was argued that the exponents may depend on more details of the
velocity statistics than only the exponents $\eta$ and
$\varepsilon$ \cite{ShS}. This idea was supported in Refs.
\cite{Falk3,shell}, where the case of short but finite correlation
time was considered for the special case of a local turnover
exponent. In those studies, the anomalous exponents were derived
to first order in small correlation time, with Kraichnan's rapid-change
model \cite{Falk3} or analogous shell model for a scalar field \cite{shell}
taken as zeroth order approximations. The exponents obtained appear
nonuniversal through the dependence on the correlation time.

In Ref. \cite{RG} and subsequent papers
\cite{RG1,cube,vektor,RG3,RG4}, the field theoretic
renormalization group (RG) and operator product expansion (OPE)
were applied to the rapid-change model and its descendants. In
that approach, anomalous scaling emerges as a consequence of the
existence in the model of composite operators with negative
scaling dimensions, identified with the anomalous exponents. This
allows one to construct a systematic perturbation expansion for
the anomalous exponents, analogous to the famous $\eps$ expansion
in the RG theory of critical behavior, and to calculate the
exponents to the second \cite{RG,RG1} and third \cite{cube}
orders. For passively advected {\it vector} fields, any
calculation of the exponents for higher-order correlations calls
for the RG techniques already in the lowest-order approximation
\cite{vektor}.

Besides the calculational efficiency, an important advantage of
the RG approach is its relative universality: it is not related to
the aforementioned solvability of the rapid-change model and can
also be applied to the case of finite correlation time or
non-Gaussian advecting field. In Ref. \cite{RG3} (see also
\cite{RG4} for the case of compressible flow) the RG and OPE were
applied to the problem of a passive scalar advected by a Gaussian
self-similar velocity with finite (and not small) correlation
time. The energy spectrum of the velocity in the inertial range
has the form $E(k)\propto k^{1-2\eps}$, while the correlation time
at the wavenumber $k$ scales as $k^{-2+\eta}$. It was shown that,
depending on the values of the exponents $\eps$ and $\eta$, the
model reveals various types of inertial-range scaling regimes with
nontrivial anomalous exponents. For $\eta>\eps$, they coincide
with the exponents of the rapid-change model and depend on the
only parameter $2\eps-\eta$, while for $\eps>\eta$ they coincide
with the exponents of the opposite (``quenched'' or ``frozen'')
case and depend only on $\eps$.

The most interesting case is $\eta=\eps$, when the exponents can
be nonuniversal through the dependence on the correlation time
(more precisely, on the ratio $u$ of the velocity correlation time
and the turnover time of the passive scalar). In the field
theoretic language, the nonuniversality of the exponents in this
regime is a consequence of the degeneracy of the corresponding
fixed point of the RG equations. It agrees with the findings of
Refs. \cite{Falk3,shell} since the borderline $\eta=\eps$,
including the ``Kolmogorov'' point $\eta=\eps=4/3$, corresponds to
the case of a local turnover exponent. It is also interesting to
note that the same relation $\eta=\eps$ for the boundary between
the time-decorrelated and quenched cases is encountered in a model
of passive advection by a strongly anisotropic flow, studied in
Refs. \cite{AvelMaj}. It was argued in \cite{AvelMaj2} that the
same boundary will be observed with very general assumptions on
the velocity statistics. Although the possibility of the
nonuniversality of anomalous exponents for $\eta=\eps$ was
demonstrated by the rigorous RG analysis, the practical
calculation by Ref. \cite{RG3} has shown that they appear
universal (independent of $u$) to the first order in $\eps$ and
$\eta$: in the one-loop approximation, the anomalous dimensions of
the relevant composite operators depend on a combination of the
model parameters (couplings) that remains constant along the line
of the fixed points. This fact is rather disappointing, because it
means that in the one-loop approximation it is impossible to judge
how the finite correlation time affects intermittency, in
particular, whether the anomalous scaling is enhanced or
suppressed in comparison with the rapid-change or quenched limits.

In this paper, we present the anomalous exponents to order
$O(\eps^{2})$ (two-loop approximation) for the most interesting
case $\eta=\eps$, including the exponents of the anisotropic
contributions, and study their dependence on $u$. [It is not
necessary to separately consider the cases $\eta>\eps$
($\eta<\eps$), because the corresponding exponents are the same as
for the rapid-change (quenched) velocity field and can be obtained
from the case $\eta=\eps$ in the limits $u\to\infty$ ($u\to0$)].

In Sec.~\ref{sec:Model}, we describe our model and its interesting
special cases. In Sec.~\ref{sec:RG}, we briefly recall the field
theoretic formulation, the RG and OPE approach to the model, and
the $O(\eps)$ result for the anomalous exponents \cite{RG3}. The
results of the two-loop calculation are presented and discussed in
Sec.~\ref{sec:Expo}: in Sec.~\ref{sec:General}, we give the
anomalous exponents to order $O(\eps^2)$ and then discuss them
separately for the isotropic (Sec.~\ref{sec:Isot}) and anisotropic
(Sec.~\ref{sec:Anis}) contributions.

The main conclusion of the paper can be formulated as follows: the
qualitative effect of the finite correlation time on the anomalous
scaling depends essentially on the correlation function
considered, the value of $u$, and the space dimensionality $d$.
For the low-order structure functions and in low dimensions ($d=2$
or 3), the inclusion of finite correlation time enhances the
intermittency in comparison with the both limits: the
time-decorrelated ($u=\infty$) and time-independent ($u=0$) ones.
Although the anomalous exponents have a well-defined limits for
$u\to0$, they show interesting irregularities in the vicinity of
the quenched limit: a rapid falloff when $u=0$ increases from
zero, with infinite slope for $d=2$, with a pronounced minimum for
$u\sim1$. On the contrary, the behavior in the region of large $u$
is smooth, like for the shell model studied in \cite{shell}. For
higher-order structure functions and large $d$, the anomalous
scaling is always weaker in comparison with the rapid-change limit
and the corresponding (positive) correction is maximal for $u=0$
and monotonically decreases to zero as $u$ tends to infinity.

\section{The model} \label{sec:Model}

The advection of a passive scalar field $\theta(x)\equiv \theta(t,{\bf x})$
is described by the stochastic equation
\begin{equation}
\nabla_{t}\theta =\nu _0\partial^{2} \theta+f, \qquad
\nabla_{t} \equiv \partial _t + v_{i}\partial_{i},
\label{1}
\end{equation}
where $\partial _t \equiv \partial /\partial t$,
$\partial _i \equiv \partial /\partial x_{i}$, $\nu _0$
is the molecular diffusivity coefficient, $\partial^{2}$ is the Laplace
operator, ${\bf v}(x)\equiv \{v_{i}(x)\}$ is the divergence-free (owing
to the incompressibility) velocity field, and $f\equiv f(x)$ is an
artificial Gaussian random noise with zero mean and correlation function
\begin{equation}
\langle  f(x)  f(x')\rangle =  C(t-t',\, {\bf r}), \qquad
{\bf r}={\bf x}-{\bf x'}.
\label{2}
\end{equation}
The form of the correlator is unessential; it is only important
that the function $C$ in Eq. (\ref{2}) decreases rapidly for $r\gg L$,
where $L$ is some integral scale. The noise maintains the steady state of
the system and, if $C$ depends on the vector ${\bf r}$ and not only its
modulus $r\equiv{\bf r}$, is a source of large-scale anisotropy. In a more
realistic formulation, the noise is replaced by an imposed constant gradient
of the scalar field; see e.g. Refs. \cite{Pumir,Siggia,RG3,RG4,OU}.

In the real problem, the velocity field satisfies the Navier--Stokes
equation. Following Refs. \cite{shell,RG3,RG4,OU}, we assume for
${\bf v}(x)$ in Eq.
(\ref{1}) a Gaussian distribution with zero mean and correlator
\begin{equation}
\langle v_{i}(x) v_{j}(x')\rangle = \int \frac{d{\bf k}}{(2\pi)^d}\,
P_{ij}(\k) \, D_{v}(t-t',k) \exp [{\rm i}{\bf k}\cdot({\bf x}-{\bf x'})],
\label{3}
\end{equation}
where $P_{ij}(\k)\equiv \delta _{ij} - k_i k_j / k^2$ is the transverse
projector and the function $D_{v}(t,k)$ will be chosen in the form
\begin{equation}
D_{v}(t-t',k) = \frac {D_{0}}{2u_{0}}\, \frac{1}{k^{d-2+2\eps}}\,
\exp [-\omega(k)(t-t')], \quad \omega(k) = u_{0}\nu_0\, k^{2-\eta}.
\label{4}
\end{equation}
Here $D_0$ and $u_0$ are positive amplitude factors and the positive
exponents $\eps$ and $\eta$ play the part of small expansion parameters
in the RG theory; see Refs. \cite{RG3,RG4}. It is also convenient to
introduce the ``coupling constant'' $g_0\equiv D_0/\nu_0^2$ (expansion
parameter in the ordinary perturbation theory). The infrared (IR)
regularization is provided by the sharp cutoff in all momentum intergals
from below at $k=m$ with $m\sim 1/L$.

The model (\ref{3}), (\ref{4}) contains two special cases that possess some
interest on their own:
in the limit $u_{0}\to\infty$, $g_{0}'\equiv g_{0}/u_{0}^{2}= {\rm const}$
we arrive at the rapid-change model,
\begin{equation}
D_{v}(\omega,k)\to g_{0}'\nu_0\, \delta(t-t')\, k^{-d-2\eps+\eta},
\label{RC1}
\end{equation}
while the limit $u_{0}\to 0$, $g_{0}''\equiv g_{0}/2u_{0}={\rm
const}$ corresponds to the case of a quenched (time-independent)
velocity field,
\begin{equation}
D_{v}(\omega,k)\to g_{0}''\nu_0^{2}\,k^{-d+2-2\eps}.
\label{RC2}
\end{equation}
The latter case has a close formal resemblance with the well-known models of
random walks in random environment with long-range correlations; see Refs.
\cite{walks1,walks3}.

\section{Renormalization group and operator product expansion}
\label{sec:RG}

The RG theory of the model (\ref{1})--(\ref{4}) is presented in Refs.
\cite{RG3,RG4} in detail; below we briefly recall only the necessary
information. The stochastic problem (\ref{1})--(\ref{4}) can be cast as
a field theory with action functional
\begin{equation}
S(\theta,\theta',v)= -{\bf v} D_{v}^{-1} {\bf v}/2 +
\theta' D_{f}\theta' /2+
\theta' \left[ - \nabla_{t} + \nu _0\partial^{2}  \right] \theta,
\label{action}
\end{equation}
where $\theta'$ is an auxiliary scalar field and $D_{f}$ and $D_{v}$ are
correlators (\ref{2}) and (\ref{3}), respectively. In Eq. (\ref{action}),
all the required integrations over $x=(t,{\bf x})$ and summations over
the vector indices are understood.

The model (\ref{action}) is logarithmic for $\eps=\eta=0$; the
ultraviolet (UV) singularities have the form of poles in various
linear combinations of $\eps$ and $\eta$ in the correlation functions.
They can be removed by the only counterterm of the form $\theta'
\partial^{2} \theta$, which is equivalent to the following multiplicative
renormalization of the parameters $g_{0}$, $u_{0}$ and $\nu_0$
in the action functional (\ref{action}):
\begin{equation}
\nu_{0}=\nu Z_{\nu}, \qquad g_{0}=g\mu^{2\eps+\eta}\, Z_{g},
\qquad u_{0}=u\mu^{\eta}\, Z_{u},
\label{mult}
\end{equation}
where $g$, $u$, and $\nu$ are the renormalized counterparts of the bare
parameters, $\mu$ is the reference mass in the minimal subtraction (MS)
scheme, which we always use in practical calculations, and
$Z_{i}=Z_{i}(g,u;d;\eps,\eta)$ are the renormalization
constants satisfying the identities
\begin{equation}
Z_{g}= Z_{\nu}^{-3},\qquad Z_{u}= Z_{\nu}^{-1}.
\label{svaz}
\end{equation}
Fixed points of the corresponding RG equations are found from the
requirement that the $\beta$ functions
\begin{equation}
\beta_{g}\equiv\Dm  g=g[-2\eps-\eta+3\gamma_{\nu}], \qquad
\beta_{u}\equiv\Dm  u=u[-\eta+\gamma_{\nu}], \qquad
\gamma_{\nu}\equiv \Dm \ln Z_{\nu}
\label{beta}
\end{equation}
vanish. Here $\Dm$ is the operation $\mu\partial_{\mu}$ for fixed
bare parameters and the relations between $\beta$ functions and
the anomalous dimension $\gamma_{\nu}$ result from the definitions
and the relation (\ref{svaz}).

From Eq. (\ref{beta}) the exact relation $\beta_{g}/g-3\beta_{u}/u
=2(\eta-\eps)$ follows which shows that the $\beta$ functions
cannot vanish simultaneously for finite values of their arguments,
except for the case $\eta=\eps$. Therefore, to find the fixed
points we must set either $u=\infty$ or $u=0$ and simultaneously
rescale $g$ so that the anomalous dimension $\gamma_{\nu}$ remain
finite. These two options correspond to the two limits (\ref{RC1})
and (\ref{RC2}), so that the rapid-change and quenched cases are
fixed points of the general model. The analysis shows that the
former is IR stable (and thus describes the inertial-range
asymptotic behavior) for $\eta>\eps$ while the latter is stable
for $\eta<\eps$.

Most interesting is the case $\eta=\eps$, when the $\beta$
functions become proportional, and the set $\beta_{g}=\beta_{u}=0$
reduces to a single equation. As a result, the corresponding fixed
point is degenerate: rather than a point, one obtains a line of
fixed points in the $g$--$u$ plane. They can be labelled by the
value of the parameter $u$, which has the meaning of the ratio of
the velocity correlation time and the scalar turnover time.

Existence of the IR stable fixed points implies certain scaling
properties of various correlation functions at scales larger than
the dissipative length $\sim g_{0}^{-1/3\eps}$. In particular, for
the equal-time structure functions
\begin{equation}
{\cal S}_{n}(\bfr)= \langle [\theta(t,{\bf x})-\theta(t,{\bf x}')]^{n}\rangle,
\qquad    \bfr = {\bf x}-{\bf x}'
\label{struc}
\end{equation}
one obtains
\begin{equation}
{\cal S}_{n}(\bfr) = D_{0}^{-n/2} r^{n(1-\eps/2)} F_{n}(m\bfr)
\label{RGR}
\end{equation}
(odd structure functions are nontrivial if the correlation function
$\langle vf \rangle$ is nonzero or if a constant gradient of the scalar field
is imposed). In the presence of anisotropy the scaling functions
$F_{n}(m\bfr)$ can be decomposed in irreducible representations of the
$SO(d)$ group. In the simplest case of uniaxial anisotropy (which is
sufficient to reveal {\it all} anomalous exponents) one can write
\begin{equation}
F_{n}(m\bfr) = P_{l} (z) F_{nl} (mr), \qquad  z = ({\bf n}\cdot {\bf r})/r,
\label{RGR1}
\end{equation}
where $P_{l}(z)$ is the $l$th order Gegenbauer polynomial (Legendre
polynomial for $d=3$) and ${\bf n}$ is a unit vector that determines the
distinguished direction.

The leading behavior of the functions $F_{nl}$ for $mr\ll0$ (inertial range)
is found from the corresponding operator product expansion and has the form
\begin{equation}
F_{nl} \propto (mr) ^{\Delta_{nl}},
\label{RGR2}
\end{equation}
where the ``anomalous exponent'' $\Delta_{nl}$ is nothing other than
the critical dimension of the irreducible traceless $l$th rank tensor
composite operator built of $n$ fields $\theta$ and minimal possible number
of derivatives \cite{RG3}. For $l\le n$ such an operator has the form
\begin{equation}
\partial_{i_{1}}\theta\cdots\partial_{i_{l}}\theta\,
(\partial_{i}\theta\partial_{i}\theta)^{p}+\cdots, \quad n=l+2p.
\label{Fnp}
\end{equation}
Here the dots stand for the appropriate subtractions involving the Kronecker
$\delta$ symbols, which ensure that the resulting expressions are traceless
with respect to any given pair of indices, for example, $\partial_{i}\theta
\partial_{j}\theta - \delta_{ij}\partial_{k}\theta\partial_{k}\theta /d$.
We also note that the numbers $n$ and $l$ necessarily have the same parity,
that is, they can only be simultaneously even or odd.

For the most interesting case of the degenerate fixed point, the dimensions
$\Delta_{nl}$ are calculated in the form of series in the only independent
exponent $\eps=\eta$, that is,
\begin{equation}
\Delta_{nl} = \sum_{k=1}^{\infty} \eps^{k} \, \Delta_{nl}^{(k)}.
\label{answer}
\end{equation}
In the lowest order one obtains \cite{RG3}:
\begin{equation}
\Delta_{nl}^{(1)} = \frac{-n(n-2)(d-1)+ \lambda_{l}(d+1)} {2(d-1)(d+2)}
\label{answer1}
\end{equation}
with $\lambda_{l}\equiv l(d+l-2)$. For $k\ge2$, the coefficients
$\Delta_{nl}^{(k)}$ depend not only on $d$ but also on the parameter $u$,
the ratio of the velocity correlation time and the scalar turnover time,
that labels fixed points in the $g$--$u$ plane (see above).

The reader not interested in the details of practical calculation can skip
the end of this section and pass to the result for $\Delta_{nl}^{(2)}$.
Calculation of the higher-order coefficients in the $\eps$ expansions for
the rapid-change model is presented in Refs. \cite{RG1,cube} in detail.
Analogous calculations for the finite correlated case are more difficult
in two respects. First, there are more relevant Feynman diagrams in the
same order of perturbation theory (for zero correlation time, many diagrams
contain closed circuits of retarded propagators and vanish). Second, and
more important distinction, is that the diagrams for the finite correlated
case involve {\it two} different dispersion laws: $\omega\propto k^{2}$ for
the scalar and $\omega\propto k^{2-\eta}$ for the velocity fields. As a
result, the calculation, as well as expressions for the renormalization
constants, become rather cumbersome already in the lowest (one-loop)
approximation; see Refs.~\cite{RG3,RG4}.

The latter difficulty can be circumvented as follows. Careful analysis shows
that in the MS scheme all the needed anomalous dimensions, $\gamma_{\nu}$
from (\ref{beta}) and $\gamma_{nl}\equiv \Dm \ln Z_{nl}$, in contrast with
the respective renormalization constants $Z_{\nu}$ and $Z_{nl}$, are
independent of the exponents $\eps$ and $\eta$ in the two-loop approximation
(for the one-loop approximation this is obvious from the explicit
expressions; see \cite{RG3,RG4}). It is thus sufficient to calculate
them for any specific choice of the exponents $\eps$ and $\eta$ that
guarantees UV finiteness of the diagrams. The most convenient choice
is $\eta=0$ and arbitrary $\eps$: all the diagrams remain finite, the
exponents in the aforementioned dispersion laws become identical, and the
practical calculations drastically simplify and become feasible.

To avoid possible misunderstandings, it should be emphasized that such an
independence is {\it not} guaranteed by the renormalizability of the model.
The renormalizability in the analytic regularization only guarantees that
the renormalization scheme can be chosen such that the correlation functions,
along with the coefficients $\beta$ and $\gamma$ in the RG equations, will
be analytic at the origin in the space of two complex variables $\eps$ and
$\eta$ \cite{Speer}. We used another scheme in which the functions $\beta$
and $\gamma$ are {\it independent} of $\eps$ and $\eta$ in the first two
orders, which does not exclude {\it nonanalytic} dependence on these
parameters in higher orders. We expect that in the three-loop approximation
nonanalytic constructions like $(\eps+\eta)/(\eps+2\eta)$ will indeed appear
in the anomalous dimensions, in particular, due to the necessity to take into
account UV finite parts of the two-loop diagrams (with our choice of the
sharp IR cutoff in Eq. (\ref{4}), the one-loop diagrams have no UV finite
parts; cf. \cite{cube} for the rapid-change case).

Thus, we conclude that the knowledge of the renormalization constants for
$\eta=0$ is sufficient to obtain the anomalous dimensions, $\beta$ functions, coordinates of the fixed points,
and the critical dimensions of composite operators for arbitrary values of
$\eta$ and $\eps$, including the most interesting case $\eta=\eps$, which we
always discuss from now on.

\section{Anomalous exponents in the two-loop approximation}
\label{sec:Expo}

\subsection{General expressions} \label{sec:General}

We have performed the complete two-loop calculation of the RG functions
(\ref{beta}) and the critical dimensions (\ref{answer}) of the composite
operators (\ref{Fnp}) for arbitrary values of $n$, $l$, $d$, and $u$ and
obtained the following expression for the second coefficient in expansion
(\ref{answer}):
\begin{eqnarray}
\Delta_{nl}^{(2)} &=& \frac{1}{(d-1)^{2}(d+2)^{2}(d+4)} \biggl\{ 2(d+4)
{\cal A} \left[ n(n-2)(d-1)+  \lambda_{l} \right] + (n-2)\Bigl\{6{\cal B}
\left[n(n-4)(d-1)+ 3\lambda_{l} \right]+
\nonumber \\
&+& 9{\cal C} \left[n(d+n)(d-1) - \lambda_{l}(d+1) \right]\Bigr\}\biggr\}
\label{answer2}
\end{eqnarray}
with $\lambda_{l}\equiv l(d+l-2)$. Here and below we denote
\begin{mathletters}
\label{ABC}
\begin{eqnarray}
{\cal A} &=& \frac{(u-1-1/u)(d+1)}{2(d+2)(1-u)} + \frac{(d+1)}
{2(d+4)(1-u)u(1+u)^{2}} F_{3}\left( \frac{1}{(u+1)^{2}}
\right)  + \frac{2ud(d+2)} {(1-d)(1-u)} {\cal J}(u,d),
\label{A}
\\
{\cal B} &=& \frac{(d+1)}{3(1-u)^{2}(d+4)} \left[\frac{u}{u+1} F_{3}
\left(\frac{1}{2(u+1)}\right) - \frac{1}{(u+1)^{2}} F_{3}
\left(\frac{1}{(u+1)^{2}}\right)
- \frac{u^{2}}{4} F_{3} \left(\frac{1}{4}\right) \right],
\label{B}
\\
{\cal C} &=& \frac{1} {9(1-u)^{2}} \biggl\{ \frac{3u^{2}(d-1)}{4}
F_{2} \left(\frac{1}{4}\right) - \frac{u[2d-1+u(d-2)]}{(u+1)} F_{2}
\left(\frac{1}{2(u+1)}\right) + \frac{[d+1+2u(d-2)]} {(u+1)^{2}} F_{2} \left(
\frac{1}{(u+1)^{2}}\right) -
\nonumber \\
&-& \frac{u^{2}(d+1)}{(d+4)} F_{3} \left(\frac{1}{4}
\right) + \frac{4u(d+1)}{(u+1)(d+4)} F_{3}\left(\frac{1}{2(u+1)}\right) -
\frac{4(d+1)} {(u+1)^{2}(d+4)} F_{3}\left(\frac{1}{(u+1)^{2}}\right)
\biggr\}.
\label{C}
\end{eqnarray}
\end{mathletters}
We also have denoted $F_{k}(x)\equiv F(1,1;d/2+k;x)$ for the hypergeometric
series
\[ F(a,b;c;z)\equiv 1+\frac{ab}{c}\, z+ \frac{a(a+1)b(b+1)}{c(c+1)}
\cdot \frac{z^{2}}{2!}+\dots \]
The values of $F_{k}$ entering into (\ref{ABC}) can be related by
the recurrent relation $(x-1)F_{2}(x)= x(d+2)F_{3}(x)/(d+4)-1$, but the
resulting expressions look more cumbersome and we shall keep both $F_{2}$
and $F_{3}$ in the formulas.

The quantity ${\cal J}(u,d)$ in Eq. (\ref{A}) can only be expressed in the
form of a single convergent integral, suitable for numerical calculation:
\begin{eqnarray}
{\cal J} (u,d) &=&
{\displaystyle \frac{\Gamma(d/2)}{\sqrt{\pi}\,\Gamma((d-1)/2)}\,
\int^1_0 dz\, \frac{(1-z^2)^{d/2}} {(u-1)^2+4uz^2}\, \Biggl\{ z^2 (1-z^2)
\ln \biggl( \frac{1+u}{2} \biggr) - z (u-1+2z^2) \arcsin z - }
\nonumber \\
&-& {\displaystyle
\frac{z(1-z^2)^{1/2}(1-u-z^2)} {[2(1+u)-z^2]^{1/2}}\, \arctan \biggl[
\frac{z [2(1+u)-z^2]^{1/2}} {(1+u-z^2)} \biggr] \Biggr\}, }
\label{J}
\end{eqnarray}
where $\Gamma(\cdots)$ is the Euler Gamma function.

The quantities (\ref{ABC}), and hence the dimensions (\ref{answer2}), have
finite limits for $u\to\infty$ and $u\to0$. In the first limit,
$\Delta_{nl}^{(2)}$ coincides with the known result for the Kraichnan's
rapid-change model (see \cite{RG} for $l=0$ and 2 and \cite{RG1} for general
$l$). The O(1/u) correction to the rapid-change limit can be found from the
following asymptotic expressions for the coefficients (\ref{ABC}):
\begin{mathletters}
\label{ABCinf}
\begin{eqnarray}
{\cal A} &=& \frac{(d+1)}{2(d+2)} \, (1+2/u) + O(1/u^{2}),
\label{Ainf}
\\
{\cal B}&=& \frac{(d+1)}{12(d+4)}F_{3}\left(\frac{1}{4}\right) \,
(1+2/u) + O(1/u^{2}),
\label{Binf}
\\
{\cal C}&=& \left[ - \frac{(d-1)}{12} F_{2}\left(\frac{1}{4}\right)+
\frac{(d+1)}{9(d+4)} F_{3}\left(\frac{1}{4}\right) \right] + \frac{1}{u}
\left[-\frac{(d-1)}{6} F_{2}\left(\frac{1}{4}\right)+
\frac{2(d+1)}{9(d+4)} F_{3}\left(\frac{1}{4}\right) +
\frac{(d-2)}{9}\right] + O(1/u^{2}).
\label{Cinf}
\end{eqnarray}
\end{mathletters}
The opposite case, $u=0$, corresponds to the quenched
(time-independent) velocity field. This case was extensively
studied in connection with the so-called ``random-random walks''
(random walks in random environments); see the review paper
\cite{walks1} and references therein. Our results for the function
$\beta_{g}$ from (\ref{beta}) and the corresponding fixed point
are in agreement with the two-loop results quoted in Ref.
\cite{walks1} for the model of random-random walks. To our
knowledge, the dimensions of composite operators (\ref{Fnp}) have
not been studied in that context, and below we give the asymptotic
expressions for the coefficients (\ref{ABC}) sufficient to find
the dimensions (\ref{answer2}) up to order $O(u)$ near the
quenched limit:
\begin{mathletters}
\label{ABCfro}
\begin{eqnarray}
{\cal A}&=& - \frac{(d+1)(3d+4)}{2d(d+2)} + u(d+1) \left\{ \frac{(3d+2)}
{2d(d-2)} - \frac{2}{d(d+2)} + \frac{\ln 2}{(d+4)} + \frac{(d+3)}{(d+4)
(d+6)} F_{4} \left(\frac{1}{2}\right) - \frac{2(d+3)}{(d+4)^{2}} \right\},
\label{A0}
\\
{\cal B}&=& - \frac{(d+1)}{3(d+2)} + \frac{u(d+1)}{3} \left\{ \frac{1}{(d+4)}
F_{3} \left(\frac{1}{2}\right) +\frac{4}{d(d+2)} \right\},
\label{B0}
\\
{\cal C}&=& \frac{(d+1)(d^{2}+4)}{9d(d+2)} + \frac{u}{9} \left\{ (1-2d)
F_{2} \left(\frac{1}{2}\right)+ \frac{4(d+1)}{(d+4)}
F_{3} \left(\frac{1}{2}\right) -\frac{4(d+1)(d+2)}{d(d-2)}
+ \frac{2(d^{2}+2d+4)}{(d+2)} \right\},
\label{C0}
\end{eqnarray}
\end{mathletters}
up to corrections of order $O(u^{2})$.

It is worth noting that the $O(u)$ terms in Eqs. (\ref{A0}) and (\ref{C0})
contain poles in $(d-2)$ and thus diverge for $d=2$. Analysis shows that,
for $d=2$, the leading correction to the $u=0$ result is not analytical in
$u$ and has the form $u\ln u$. Formally, the singularity at $d=2$ is
explained as follows. Some of the two-loop diagrams contain ``energy
denominators'' of the form $(\k+\q)^{2}+O(u)$, where $\k$ and $\q$ are
two independent integration momenta. The numerators contain factors
$\propto (P_{ij}(\k) q_{i}q_{j})^2$ stemming from transverse projectors
in the propagators. These factors suppress the singularity at $\k=-\q$,
occurring in the denominators for $u=0$, and ensure the existence of the
integrals over $\k$ and $\q$, but the ``collinear'' divergence at $\k=-\q$
occurs if the $O(u)$ to the denominators is taken into account. Physically,
this divergence can be related to a strong resonant interaction between the
excitations of the passive scalar field with the opposite momenta $\k=-\q$
of equal moduli in two dimensions. We shall see below that this singularity
remarkably affects the behavior of the dimensions (\ref{answer2}) for the
values of $d$ much larger than $d=2$.

Many studies have been devoted to the analysis of the
inertial-range turbulence in the limit $d\to\infty$
\cite{Falk1,Infty,FFR,AR}. Our model has no finite ``upper
critical dimension,'' above which anomalous scaling would vanish.
Like in the rapid-change case \cite{Infty} and, probably, in the
Navier-Stokes turbulence \cite{FFR,AR}, the anomalous scaling
disappears at $d=\infty$, but it reveals itself already in the
$O(1/d)$ approximation. Along with the results \cite{Falk1} for
the scalar rapid-change model, where the $O(1/d)$ expression for
the anomalous exponents were derived for any $\eps$, this confirms
the importance of the large-$d$ expansion for the issue of
anomalous scaling in fully developed turbulence.

Straightforward analysis of the expressions (\ref{ABC}) shows
that, for $d\to\infty$, one has ${\cal A} = O(d^0)$ [it is
important here that ${\cal J}(u,d) = O(1/d)$], ${\cal B} = O(d^0)$
and ${\cal C} = O(d)$, namely,
\begin{equation}
{\cal C} = \frac{(u+2)(3u+2)}{36(u+1)^{2}}\, d +O(d^0).
\label{CD}
\end{equation}
It then follows that for large $d$, the dimension (\ref{answer2}) behaves
as $O(1/d^2)$ and is completely determined by the only contribution with
the coefficient ${\cal C}$. This gives
\begin{equation}
\Delta_{nl}^{(2)}=\frac{(n-2)(n-l)(u+2)(3u+2)}{4(u+1)^{2}d^{2}} +O(1/d^3).
\label{DD}
\end{equation}
The general expressions (\ref{answer2}), (\ref{ABC}) are rather cumbersome,
and in the subsequent sections we shall separately discuss isotropic
contribution (even $n$, $l=0$) and anisotropic ones (general $n$, $l\ne0$).

\subsection{Isotropic sectors} \label{sec:Isot}

Expression (\ref{answer2}) simplifies for the most important case of the
isotropic sector (even $n$ and $l=0$):
\begin{eqnarray}
\Delta_{n0}^{(2)} =  \frac{n(n-2)}{(d-1)(d+2)^{2}(d+4)} \biggl\{
2(d+4) {\cal A}+ 6(n-4) {\cal B}+9(d+n) {\cal C} \biggr\}.
\label{L=0}
\end{eqnarray}
This gives $\Delta_{20}^{(2)} =0$ in agreement with the exact result
$\Delta_{20}=0$ \cite{RG3}. This means that the second-order structure
function is not anomalous. The formal proof is based on certain Schwinger
equation, which has the meaning of the energy conservation law; it is almost
identical to the analogous proof for the Kraichnan model, given in Ref.
\cite{RG}.

For the simplest nontrivial case $n=4$ one obtains
\begin{eqnarray}
\Delta_{40}^{(2)} = 8(2{\cal A}+9{\cal C})/(d-1)(d+2)^{2},
\label{L=0,N=4}
\end{eqnarray}
that is, the quantity ${\cal B}$ does not enter into the result. For
$n\ge6$, all the coefficients (\ref{ABC}) contribute to the result.

\begin{figure}
\centerline{
\hbox{
\epsfig{file=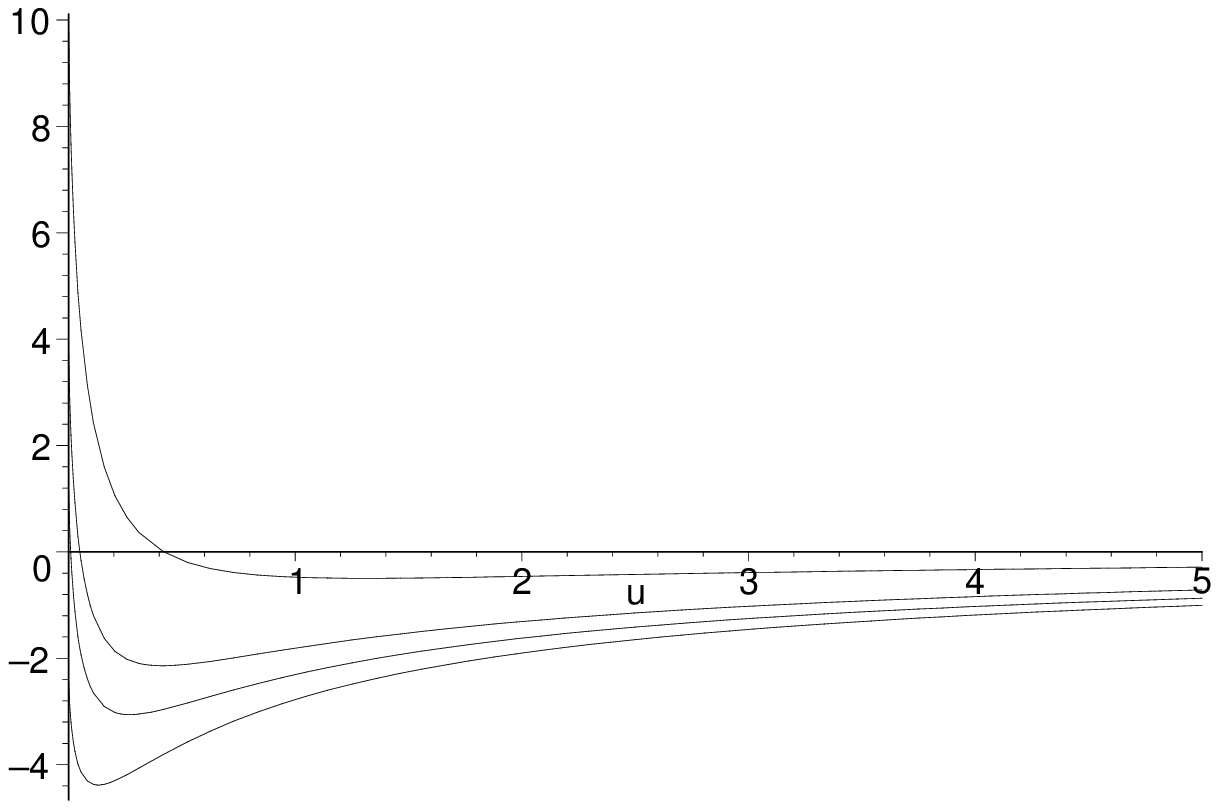,width=4cm}\hskip .5 cm
\epsfig{file=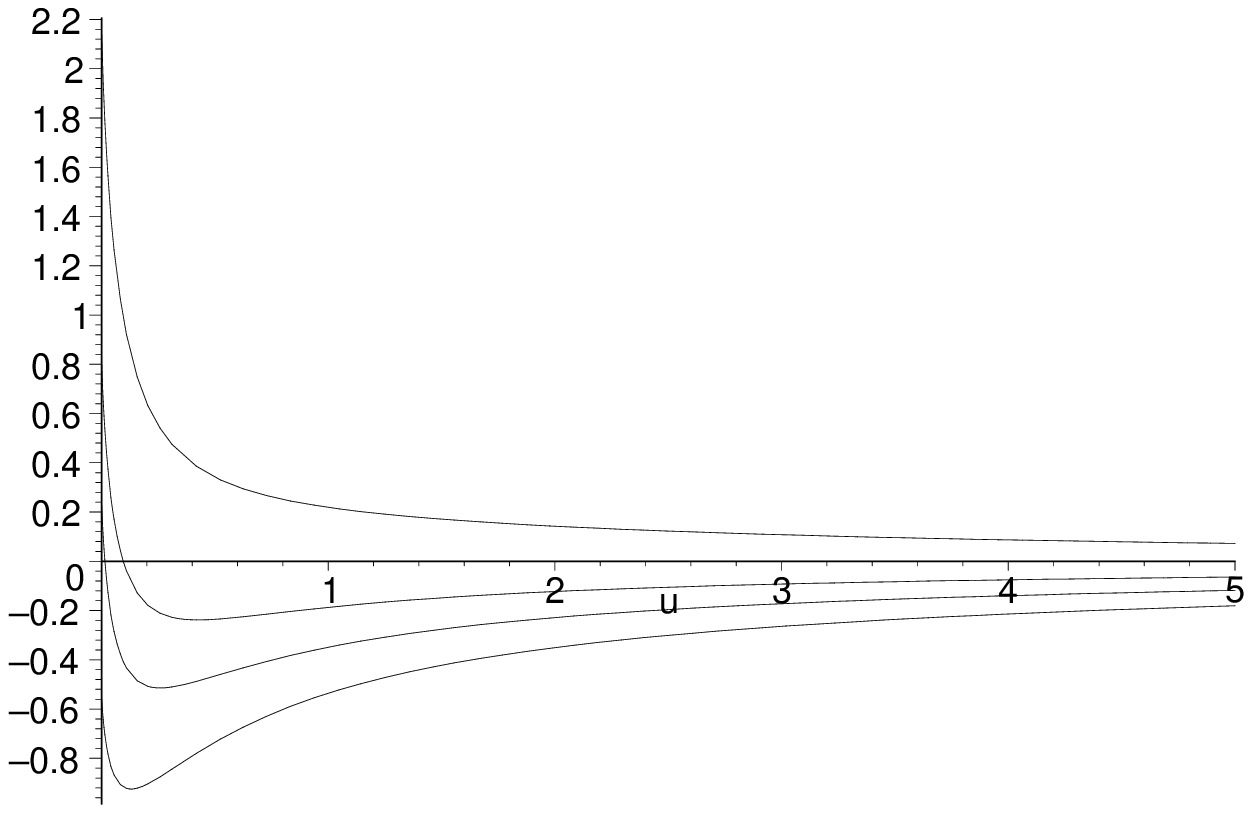,width=4cm}\hskip .5 cm
\epsfig{file=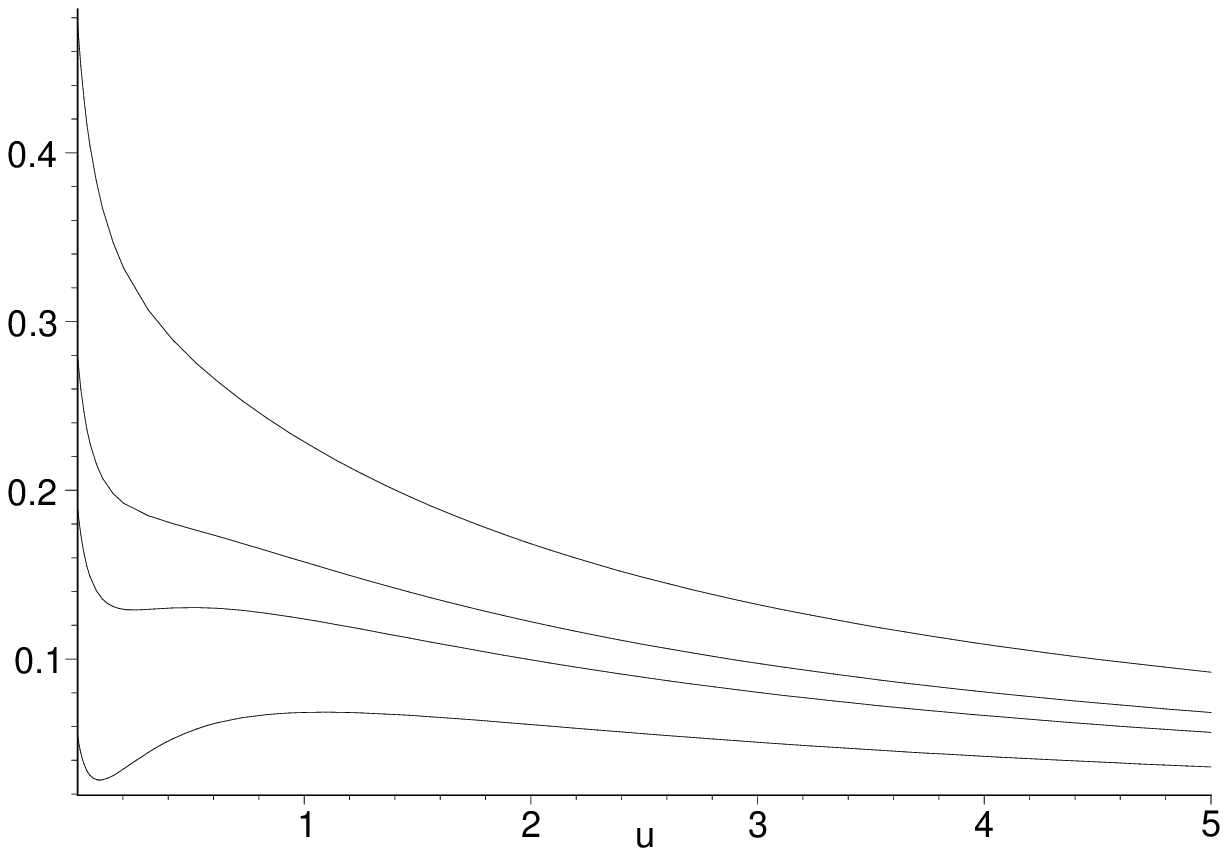,width=4cm}\hskip .5 cm
\epsfig{file=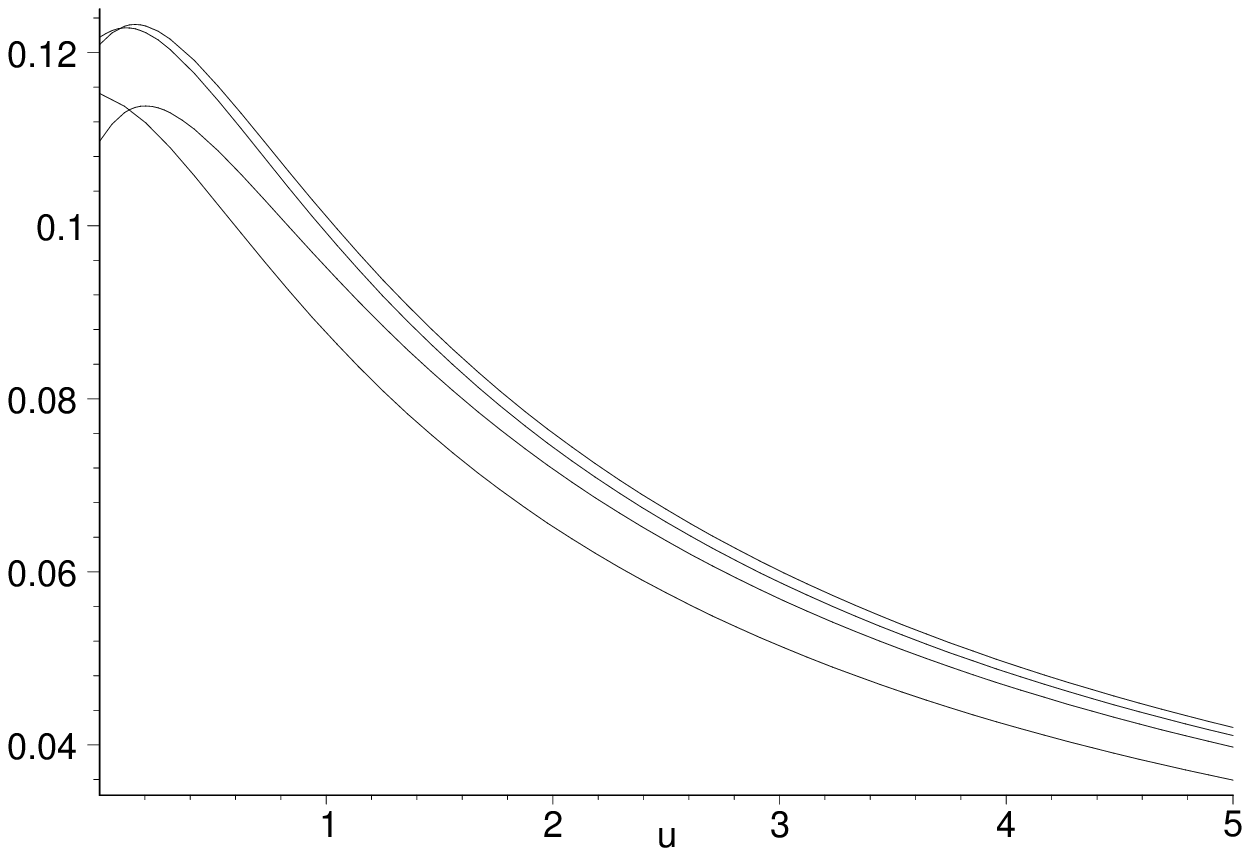,width=4cm}\hskip .5 cm } }
\vspace{0.5cm}
\caption{Behavior of the quantity $\zeta_{n}$ from Eq.~(\protect\ref{zetan})
for $n=4$, 6, 8 and 20 (from below to above) as a function of $u$ for $d=2$,
3, 5, and 10 (from the left to the right) in the units of $10^{-3}$.}
\label{Fig1}
\end{figure}

In Fig.~\ref{Fig1}, we show the behavior of the quantity
\begin{eqnarray}
\zeta_{n} \equiv
\left[\Delta_{n0}^{(2)} - \Delta_{n0}^{(2)}|_{u=\infty}\right]/n^{3}
\label{zetan}
\end{eqnarray}
for $n=4$, 6, 8 and 20 (from below to above) as a function of $u$
for several values of $d$. We have subtracted the value of the
dimension for the rapid-change case, such that the curves approach
zero as $u\to\infty$, and divided the difference by $n^{3}$, such
that the results for different $n$'s have the same order of
magnitude [the quantity (\ref{answer2}) is a third-order
polynomial in $n$]. It is worth noting that, since the leading
coefficient (\ref{answer1}) is independent of $u$, it drops from
the difference $\Delta_{n0} - \Delta_{n0}|_{u=\infty}$ of the {\it
exact} dimensions, and in the leading order $O(\eps^{2})$ the
latter is proportional to the quantity $\zeta_{n}$ introduced
above.

As one can easily see from Fig.~\ref{Fig1}, the qualitative
behavior of $\zeta_{n}$ depends essentially on the values of $n$
and $d$. For moderate $n$ and $d$ (e.g., $n=4$, 6 and 8 for $d=2$
and 3), finite correlation time enhances the intermittency
(anomalous dimensions become more negative) in comparison with the
both limits: the rapid-change ($u=\infty$) and quenched ($u=0$)
ones. While the $O(1/u)$ correction leads to a smooth decrease of
$\zeta_{n}$ for increasing $1/u$ from zero (in agreement with the
numerical simulation of Ref. \cite{shell} for a shell model near
the rapid-change limit), the rapid falloff of $\zeta_{n}$ is
observed for increasing $u$ from the quenched limit $u=0$. As a
result of the competition between these two effects, $\zeta_{n}$
is not a monotonous function of $u$ and has a pronounced minimum
in the interval between 0 and 1.

We recall that the slopes of the functions $\zeta_{n}$ at $u=0$
are infinite in two dimensions for all values of $n$ due to the
presence of poles $(d-2)$ in the $O(u)$ terms in Eqs. (\ref{A0})
and (\ref{C0}); see Sec.~\ref{sec:General}. For $d>2$, the slopes
become finite but they still remain very steep for $d=3$ and lead
to a rapid falloff of $\zeta_{n}$, as Fig.~\ref{Fig1}(b) shows.
This fact also suggests that the quenched case, in contrast with
the rapid-change one, can hardly serve as a good zero-order
approximation in studying more realistic models of passive
advection by the velocity field with finite correlation time.

If $n$ or $d$ is large enough, the minimum becomes less
pronounced, the behavior of $\zeta_{n}$ becomes more regular
(e.g., $n=20$ for $d=2$) and eventually $\zeta_{n}$ becomes a
monotonically decreasing function of $u$ ($n=20$ for $d=3$). For
such values of $n$ or $d$, the function $\zeta_{n}$ approaches
zero at $u\to\infty$ from above. In other words, the $O(1/u)$
correction to the rapid-change limit suppresses the intermittency,
in contrast with the case of moderate $n$.

In the limit of large $d$, from Eq. \ref{DD} one easily obtains
\begin{equation}
\Delta_{n0}^{(2)} - \Delta_{n0}^{(2)} |_{u=\infty}
= \frac{n(n-2)(2u+1)} {4(u+1)^{2}d^{2}} +O(1/d^{3})
\label{DD2}
\end{equation}
(note that the $O(n^{3})$ term in this approximation disappears and
$\Delta_{n0}^{(2)}$ becomes only quadratic in $n$). One can see that for
all values of $n$, the difference (\ref{DD2}) is positive, decreases
monotonically when $u$ grows, and approaches zero from above when
$u\to\infty$.

In Ref. \cite{Falk3}, the $O(1/u)$ correction to the rapid-change case
was derived by the zero-mode techniques in the limit of large $d$ and
arbitrary (not small) values of $\eps$, for the case of a local turnover
exponent ($\eps=\eta$). Although the anomalous exponents were shown
to be nonuniversal (dependent on $u$), our results disagree with Ref.
\cite{Falk3} in two respects. First, due to the universality (independence
of $u$) of the leading term (\ref{answer1}), the ratio (\ref{DD2}) is of
order $O(\eps^{2})$ and not $O(\eps)$. Second, the $O(1/u)$ correction in
(\ref{DD2}) is positive for all $n$, while, according to \cite{Falk3},
inclusion of the finite correlation time makes the anomalous exponents more
negative in comparison with the rapid-change limit for all $n$ and $\eps$.
It is not clear whether this disagreement can be explained by some
distinctions between our model (\ref{3}), (\ref{4}) and the velocity
ensemble employed in \cite{Falk3}. It is possible to show, however, that
any modification of the function (\ref{4}) consistent with the RG analysis
performed in Ref. \cite{RG3} and section \ref{sec:RG} above, leads to an
universal (independent of $u$) expression for the leading term in
$\Delta_{nl}^{(1)}$, so that the $O(1/u)$ correction to the zero-correlated
limit remains of order $O(\eps^{2})$.

The changeover from the behavior typical to low spatial dimensions to the
behavior described by Eq. (\ref{DD2}) also produces interesting patterns,
as illustrated by Fig.~\ref{Fig1} for $d=5$ and~10.

\subsection{Anisotropic sectors} \label{sec:Anis}

Let us turn to the analysis of anisotropic contributions in the structure
functions (\ref{RGR}), (\ref{RGR1}), described by the dimensions
(\ref{answer}) with $l\ne0$. We recall that such contributions appear
in the inertial-range expression (\ref{RGR1}) if the forcing (\ref{2}) is
chosen to be anisotropic, or a constant gradient of the scalar field is
imposed.

An important property of the first-order result (\ref{answer1}) is that for
any fixed $n$, the quantity $\Delta_{nl}^{(1)}$ increases monotonically with
$l$ \cite{RG1}. One can say that the exponents, associated with tensor
composite operators (\ref{Fnp}), exhibit a kind of hierarchy related to the
degree of anisotropy: the less is the rank $l$, the less is the dimension
and, consequently, the more important is the corresponding contribution to
the inertial-range expression (\ref{RGR1}). The leading terms in the even
structure functions (\ref{struc}) are given by the scalar operators
(\ref{Fnp}) with $l=0$, that is, they are the same as in the model with
isotropic forcing (we recall that $n$ and $l$ should be simultaneously
even or odd).

This behavior is in agreement with the existing phenomenological ideas,
according to which the anisotropy introduced at large scales by the forcing
(boundary conditions, geometry of an obstacle {\it etc.}) dies out when the
energy is transferred down to smaller scales owing to the cascade mechanism
\cite{Legacy}. The hierarchy of anisotropic contributions appears rather
universal, being also observed for a vector (magnetic) field, advected by
the Kraichnan velocity ensemble \cite{Lanotte2}, for the scalar advected by
the two-dimensional Navier--Stokes velocity field \cite{CLMV99} and for the
turbulent velocity field itself \cite{Arad1}.

Nevertheless, the anisotropy survives in the inertial range and reveals
itself in dimensionless ratios involving {\it odd} structure functions,
\begin{equation}
{\cal R}_{k} \equiv {\cal S}_{2k+1} / {\cal S}_{2}^{k+1/2}.
\label{skew}
\end{equation}
For a number of models it was shown that the skewness factor ${\cal R}_{1}$
decreases down the scales but slower than predicted by phenomenological
theories \cite{Pumir,Siggia}, while the higher-order odd ratios
(hyperskewness ${\cal R}_{2}$ {\it etc.}) increase, thus signalling of
persistent small-scale anisotropy \cite{RG3,RG4,Lanotte2,CLMV99}.
Due to the aforementioned hierarchy of the dimensions (\ref{answer1}), the
leading terms in the odd structure functions (\ref{struc}) in our model are
determined by the vector operators (\ref{Fnp}) with $l=1$, and it is easy
to check the above statements from the explicit expression (\ref{answer1}).

Of course, the $O(\eps^{2})$ contribution (\ref{answer2}) cannot change all
the above properties of the dimensions $\Delta_{nl}$, determined by their
leading $O(\eps)$ contribution (\ref{answer1}), as far as small values of
$\eps$ are concerned. However, since the dependence on $u$ occurs only in
the $O(\eps^{2})$ contribution, it should be taken into account if one
wishes to discuss how finite correlation time affects the hierarchy of
the dimensions or the behavior of the dimensionless ratios.

Consider the effects of the finite correlation time on the hierarchy of the
anisotropic contributions. To this aim, consider the difference of the
coefficients (\ref{answer2}) for a fixed value of $n$ and two neighboring
subsequent values of $l$:
\begin{eqnarray}
\Delta_{n,l+2}^{(2)}-\Delta_{nl}^{(2)} = \frac{2(2l+d)\,
{\cal K}_{n} (d,u)} {(d-1)^{2}(d+2)^{2}(d+4)} \, , \qquad
{\cal K}_{n} (d,u) \equiv \Bigl\{ 2(d+4) {\cal A}+ 9(n-2)
\bigl[ 2 {\cal B}-(d+1) {\cal C} \bigr] \Bigr\}.
\label{L}
\end{eqnarray}
(we recall that for a fixed $n$, all possible values of $l$ are either
even or odd, so that the subsequent values of $l$ differ by 2). It is clear
from Eq. (\ref{L}) that the sign and the dependence on $u$ of the whole
expression is determined by the behavior of the function ${\cal K}_{n}(d,u)$.

\begin{figure}
\centerline{
\hbox{
\epsfig{file=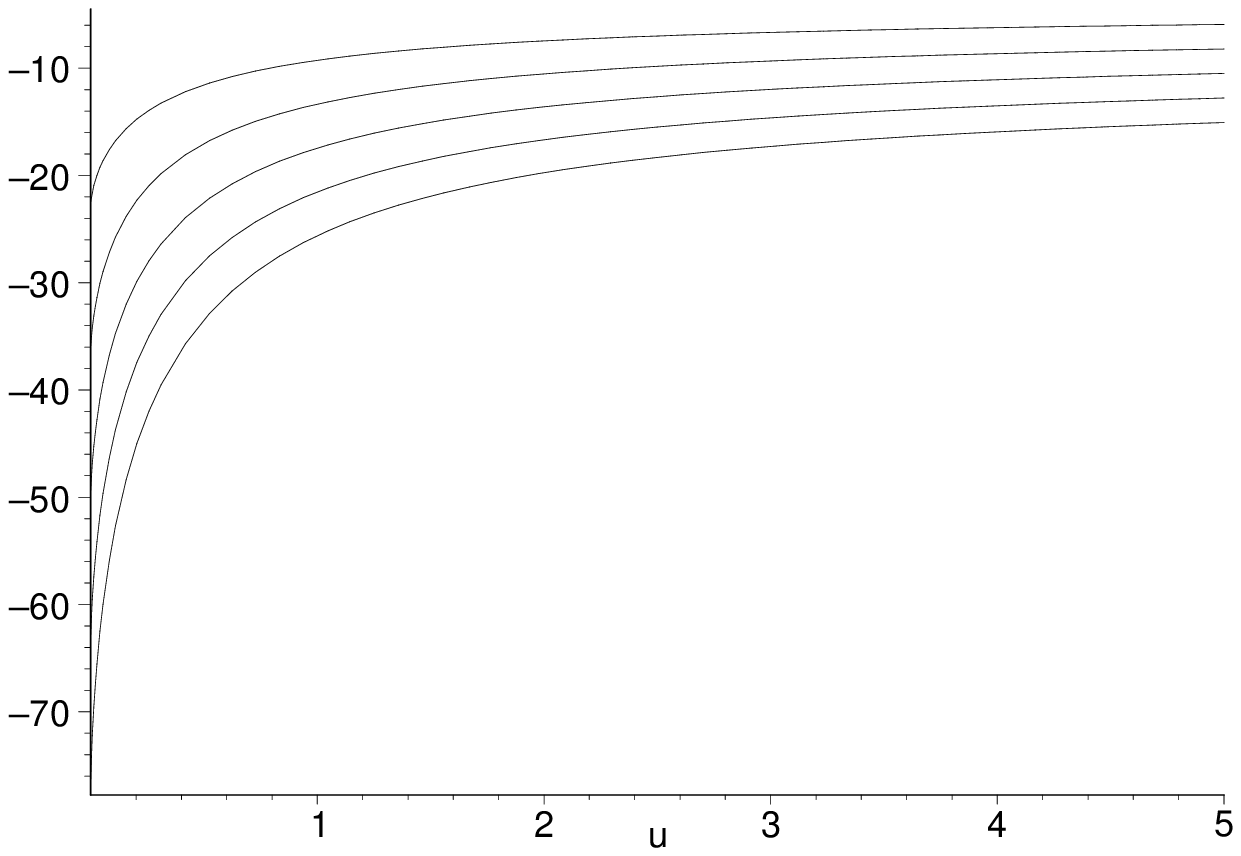,width=7cm}\hskip .5 cm
\epsfig{file=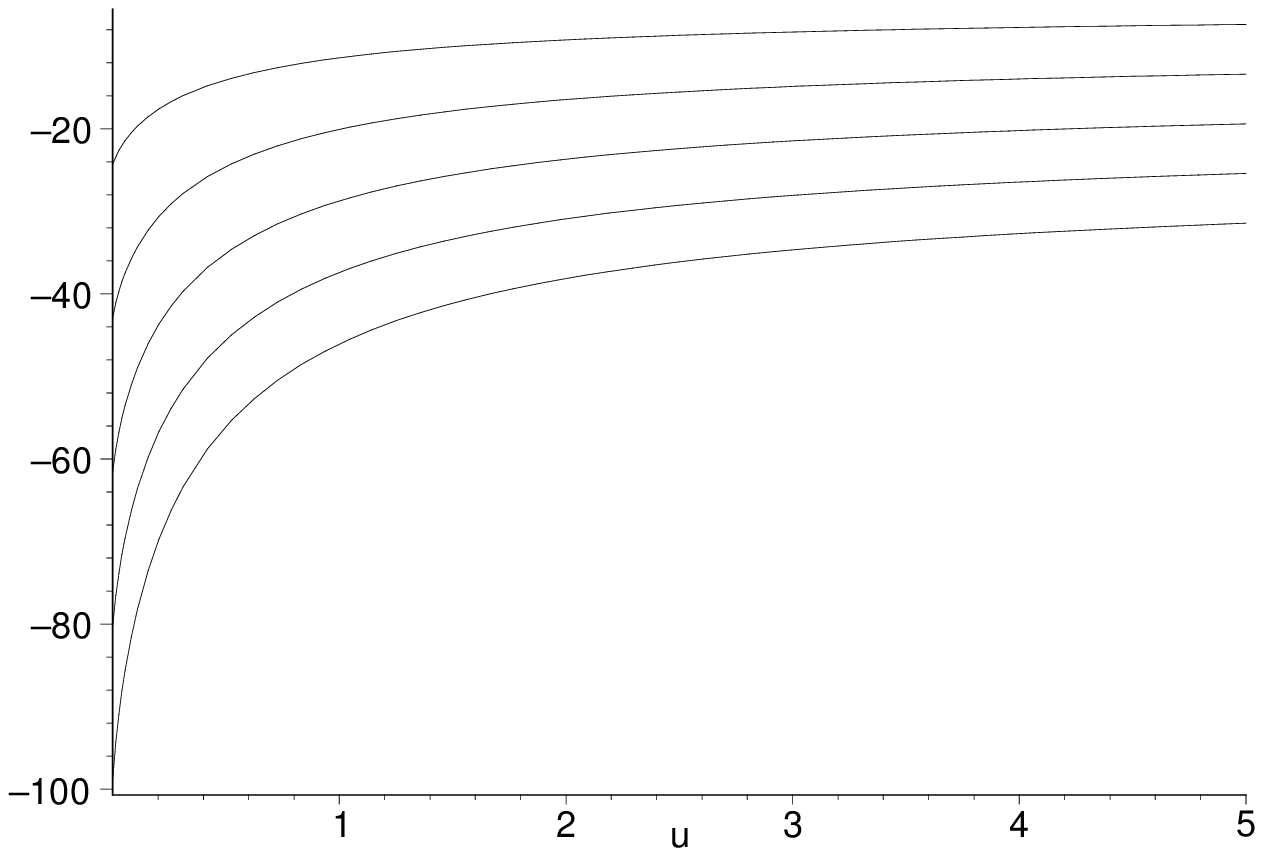,width=7cm}\hskip .5 cm } }
\vspace{0.5cm}
\caption{Behavior of the quantity ${\cal K}_{n}(d,u)$ from
Eq.~(\protect\ref{L}) as a function of $u$ for $n=2$, 3, 4, 5, and 6
(from above to below) for $d=2$ (left) and $d=3$ (right).}
\label{Fig3}
\end{figure}

In Fig.~\ref{Fig3}, we plot the quantity ${\cal K}_{n}(d,u)$ as a function
of $u$ for $n=2$, 4, 6 and 20 (from above to below) for $d=2$ (a) and
$d=3$ (b). The function is always negative for all the cases studied and
increases monotonically with $u$. This behavior persists in the limit of
large $d$, as follows from the asymptotical expression (\ref{DD}).

We thus conclude that the $O(\eps^{2})$ contribution in the exact dimension
(\ref{answer}) ``tries to cope'' with the hierarchy, set by the $O(\eps)$
term, for all values of $n$, $l$, $d$ and $u$; this effect is at its
strongest for $u=0$ and weakens monotonically as $u$ increases from 0
to $\infty$.

Now let us turn to the dimensionless ratios ${\cal R}_{k}$ in
(\ref{skew}). From the discussion below Eq. (\ref{skew}) and
asymptotic representation (\ref{RGR}) one obtains the power-like
inertial-range asymptotic expression ${\cal R}_{k}\propto
(mr)^{\Delta_{2k+1,1}}$ with $\Delta_{2k+1,1}$ from Eq.
(\ref{answer}) [we recall that in our model $\Delta_{2,0}=0$; see
Sec.~\ref{sec:Isot}]. Due to the $u$-independence of the
first-order answer (\ref{answer1}), the $O(\eps)$ contribution in
the exponent $\Delta_{2k+1,1} = \eps (d+2-4k^2) /2(d+2) +
O(\eps^{2})$ coincides with its analog for the Kraichnan model;
see Refs. \cite{Pumir} for $k=1$ and \cite{RG4} for general $k$.
It completely determines the qualitative behavior of the
quantities ${\cal R}_{k}$: for $k=1$ one has $\Delta_{3,1}>0$ and
the skewness factor ${\cal R}_{1}$ decreases with $mr$, while for
$k\ge1$ one has $\Delta_{2k+1,1}<0$ and the higher-order ratios
${\cal R}_{k}$ increase for $mr\to0$.

\begin{figure}
\centerline{
\hbox{
\epsfig{file=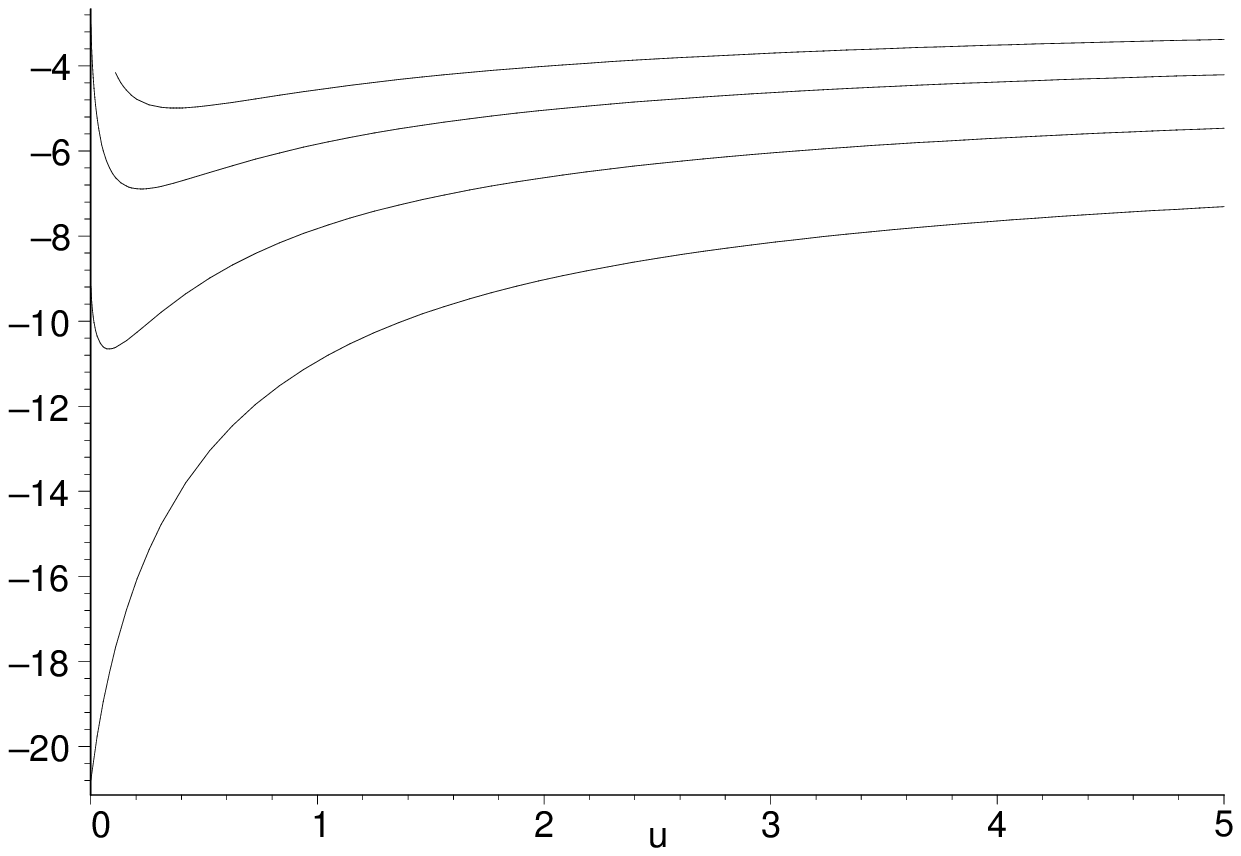,width=7cm}\hskip .5 cm
\epsfig{file=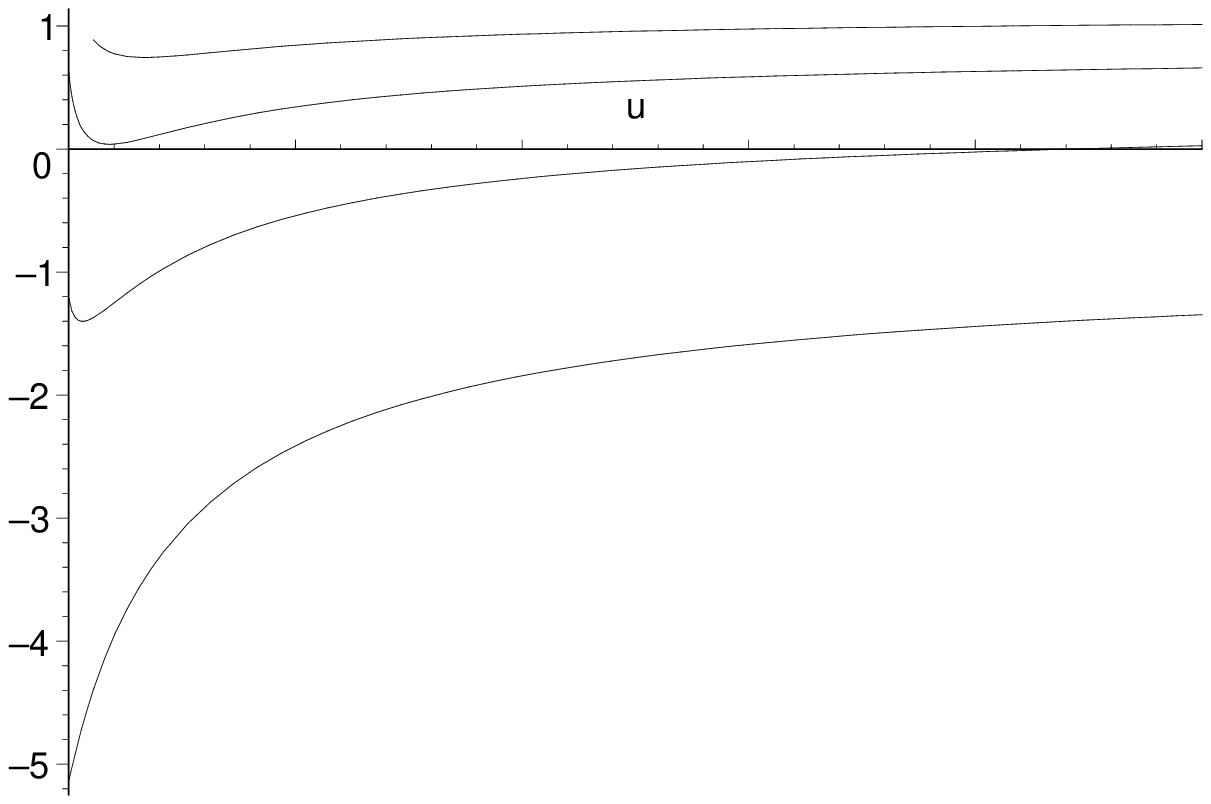,width=7cm}\hskip .5 cm } }
\vspace{0.5cm}
\caption{Behavior of the quantity
$\xi_k\equiv\Delta_{2k+1,1}^{(2)}/(2k+1)^3$ from
Eq.~(\protect\ref{answer2}) as a function of $u$ for $k=1$, 2, 3 and 4,
(from below to above) for $d=2$ (left) and $d=3$ (right), in the units of
$10^{-3}$.}
\label{Fig4}
\end{figure}

In Fig.~\ref{Fig4}, we show the behavior of the second-order
correction $\Delta_{2k+1,1}^{(2)}$, obtained from the general
formula Eq. (\ref{answer2}) and divided by $(2k+1)^{3}$ like for
the even dimensions, as a function of $u$ for $k=1$, 2, 3 and 4
(from below to above) for $d=2$ (a) and $d=3$ (b).

One can see that the effect of the $O(\eps^2)$ correction on the
inertial-range behavior of the ratios ${\cal R}_{k}$ is different
for different $d$, $k$ and $u$. In two dimensions, the corrections
are negative for all $u$ and moderate $k$: the decay of the
skewness factor ${\cal R}_{1}$ for $mr\to0$ appears even slower
than indicated by the $O(\eps)$ approximation, while the growth of
the ratios with $k\ge2$ becomes faster.

In three dimensions, the correction is negative for $k=1$ so that
the decay ${\cal R}_{1}$ for $mr\to0$ is also slower than in the
$O(\eps)$ approximation. For $k=2$, the correction is negative for
small $u$ (so that the growth of the hyperskewness factor ${\cal
R}_{2}$ for $mr\to0$ is faster than in the first-order
approximation), but it changes its sign for some finite value of
$u$ and the growth of ${\cal R}_{2}$ slows down. For $k\ge3$, the
corrections are negative for all $u$ and the growth of the
corresponding higher-order ratios ${\cal R}_{k}$ appears slower
than predicted by the $O(\eps)$ expression. One thus may conclude
that for $d=3$, with the exception of the $k=2$ case, the effect
of the second-order term is opposite to the tendency set by the
first-order approximation.

For $k$ large enough and any $d$, the behavior of the quantities
$\Delta_{2k+1,1}^{(2)}$ becomes similar to that of the even
dimensions $\Delta_{2k,0}^{(2)}$ discussed in Sec.~\ref{sec:Isot}:
they decrease monotonically as $u$ increases, comparatively fast
for small $u$ (due to the singularity in the slope for $d=2$; see
Sec.~\ref{sec:General}) and rather slow when $u$ becomes large
enough. This follows from the fact that the $l$-independent
contribution in the general expression $\Delta_{nl}^{(2)}$ behaves
as $O(n^3)$ for $n\to\infty$, while its $l$-dependent contribution
behaves only as $O(n)$; see Eq.~(\ref{answer2}).

We also note that for moderate $k$, the quantities
$\Delta_{2k+1,1}^{(2)}$ show a nonmonotonous dependence on $u$ in
the region of small $u$ and in this respect they also resemble the
even dimensions; see Fig.~\ref{Fig1} and the discussion in
Sec.~\ref{sec:Isot}.

\section{Conclusion} \label{sec:Con}

We have applied the RG and OPE methods to a simple model of a
passive scalar quantity advected by the synthetic Gaussian
velocity field with a given self-similar covariance with finite
correlation time. The structure functions of the scalar field
exhibit inertial-range anomalous scaling behavior, as a
consequence of the existence in the model of composite operators
with negative scaling dimensions, identified with anomalous
exponents.

For the special case of a local turnover exponent, the anomalous
exponents are nonuniversal through the dependence on a
dimensionless parameter $u$ that has the meaning of the ratio of
the velocity correlation time and the scalar turnover time. The
universality reveals itself only in the second order of the $\eps$
expansion, and we have derived the exponents to order
$O(\eps^{2})$, including anisotropic contributions.

It is shown that, for isotropic contributions, the qualitative
effect of finite correlation time depends essentially on $n$, the
order of the structure function, and the space dimensionality $d$.
For moderate $n$ and $d$, finite correlation time enhances the
intermittency in comparison with the both limits: the rapid-change
($u=\infty$) and quenched ($u=0$) ones. The $O(\eps^{2})$ term
shows a highly nontrivial behavior in the vicinity of the quenched
limit: a rapid falloff when $u=0$ increases from zero, with
infinite derivative at $u=0$ for $d=2$, with a pronounced minimum
for $u\sim1$. This irregularity shows that the time-independent
advecting field can hardly be a reasonable approximation in
studying more realistic models of passive advection by the
velocity field with finite correlation time. The behavior near the
opposite limit, $u=\infty$, is smooth in agreement with the
existing simulation for a shell model \cite{shell}.

The behavior changes remarkably when $n$ and/or $d$ become large
enough: the correction to the limit $u=\infty$ due to finite
correlation time is positive for all $u$ (that is, the anomalous
scaling is suppressed in comparison with the rapid-change case),
it is maximal for $u=0$ and monotonically decreases to zero as $u$
tends to infinity.

In the anisotropic sectors, the $O(\eps^2)$ terms diminish the
hierarchy revealed by the first-order terms for all values of the
parameters $n$, $l$ and $d$; this effect is maximal at $u=0$ and
decreases monotonically with $1/u$.

The effect of the $O(\eps^2)$ corrections on the inertial-range
behavior of the dimensionless ratios involving odd-order structure
functions depends on $d$. For $d=2$ and moderate $k$ these
corrections are negative; the decay of the skewness factor ${\cal
R}_{1}$ for $mr\to0$ is slower while the growth of the
higher-order ratios ${\cal R}_{k}$ with $k\ge2$ is faster than
indicated by the $O(\eps)$ approximation by Refs.
\cite{Pumir,RG3}. For $d=3$, the effect is, for most cases,
opposite to the tendency set by the first-order approximation: the
decay of the skewness factor slows down as well as the growth of
the higher-order ratios.

Our analysis has been confined with the region of small $\eps$,
where the results obtained within the $\eps$ expansion are
internally consistent and undoubtedly reliable [we recall again
that, although the leading terms of the anomalous exponents are of
order $O(\eps)$, the leading terms in which the effects of finite
correlation time occur are of order $O(\eps^2)$]. We do not
discuss here the serious issue of validity of the $\eps$
expansions for finite $\eps=O(1)$. One can think that, in our
model, the natural region of validity of the $\eps$ expansion is
restricted by the value $\eps=1/2$, where the velocity field
acquires negative critical dimension (along with all its powers)
and new IR singularities, related to the well-known sweeping
effects, occur in the diagrams; see the discussion in Ref.
\cite{RG3}. (It should be noted, however, that such singularities
do not necessarily lead to a changeover in the inertial-range
behavior, as shown in \cite{RG3} for the special case of the
structure function ${\cal S}_{2}$ for $u=0$). On the other hand,
$\eps=1/2$ can be regarded as the upper bound of the range of
validity of the model itself: the lack of Galilean covariance
becomes a serious drawback of the synthetic Gaussian velocity
ensemble when the sweeping effects become important. The next
important step should be the analytical derivation of anomalous
exponents of a passive scalar advected by the Galilean covariant
velocity field; this work is now in progress.

\acknowledgments

The authors are thankful M.~Hnatich, A.~Kupiainen, P.~Muratore Ginanneschi,
M.~Yu. Nalimov, A.~N. Vasil'ev, and A.~Vulpiani for discussions. The
work was supported by the Nordic Grant for Network Cooperation
with the Baltic Countries and Northwest Russia No.~FIN-18/2001.
N.V.A. and L.Ts.A. were also supported by the program
``Universities of Russia'' and the GRACENAS Grant No.~E00-3-24.
N.V.A. and L.Ts.A. acknowledge the Department of Physical Sciences
of the University of Helsinki for kind hospitality.

\end{document}